\documentclass[twocolumn,showpacs,amsmath,amssymb,prc]{revtex4}

\usepackage{graphicx}      
\usepackage{dcolumn}       
\usepackage{bm}            

\begin{document}

\title{Regularization of Legendre Function Series for
Charged Particles
\\
Improved Nearside-Farside Subamplitudes}

\author{R. Anni}
\affiliation{Dipartimento di Fisica dell'Universit\`a and
Istituto Nazionale di Fisica Nucleare, I73100 Lecce, Italy}

\date{August 20, 2002} 

\begin{abstract}
A simple regularization procedure is proposed for the
Legendre function series of improved nearside-farside 
subamplitudes for charged particles elastic scattering.
The procedure is the extension of the usual one which 
defines the partial wave series  for the scattering 
amplitude in the presence of a long range Coulomb term in 
the potential, and it provides the same convergence rate.
\end{abstract}


\maketitle

The nearside-farside (NF) method proposed by Fuller \cite{FUL75} 
is an effective tool to separate the full elastic scattering amplitude 
$f(\theta)$, where $\theta$ is the scattering angle,
into simpler subamplitudes \cite{HUS84,BRA97}.
The Fuller NF subamplitudes are usually 
more slowly varying and less structured then $f(\theta)$.
This allows one to explain the complicated patterns appearing 
in some cross sections, $\sigma(\theta)=|f(\theta)|^2$, as interference 
effects between simpler nearside (N) and farside (F) subamplitudes.
These subamplitudes can often be interpreted as 
contributions from simple scattering mechanisms allowing a 
physical understanding of the scattering process \cite{HUS84}.

Sometimes, particularly when applied to scattering 
of $\alpha$ particles and light heavy-ions at intermediate 
and high energies, the Fuller NF subamplitudes are biased
by the presence of unphysical contributions, making the 
NF subamplitudes more structured 
then desired.
Recently an improved NF method has been proposed 
\cite{ANN01d,ANN02b} to further extend the effectiveness of 
the original Fuller technique.
The improved NF method is based on 
a modified \cite{WHI01} Yennie, Ravenall, and Wilson (YRW) 
\cite{YEN54} resummation identity,  
which holds for Legendre polynomial series (LPS).
The increased effectiveness descends from  using 
resummation parameters with values reducing the unphysical 
contributions to the Fuller NF subamplitudes.

The Legendre function series (LFS) for the improved NF 
subamplitudes are, however, not convergent in the usual sense.
A resummation technique \cite{ANN81}, named in the following 
extended YRW (EYRW) resummation,
was used in Refs. \cite{ANN01d,ANN02b} to obtain
convergent series.
At forward angles, the 
rate of convergence of the EYRW series is not 
satisfactory in the presence of a long range Coulomb term in the potential.
For $\alpha$ particles, light and heavy ions scattering this 
fact is disturbing, because it compels one to use more 
partial waves then necessary in standard optical potentials
calculations and in the usual Fuller NF method.
Here we present a regularization procedure that, if
applied to LFS of improved NF subamplitudes,
makes these series as rapidly convergent as 
those of more conventional approaches.

The starting point for the improved NF method is 
the quantum mechanical partial wave series (PWS) of the 
elastic scattering amplitude 
\begin{equation}
f(\theta )=\sum_{l=0}^{\infty }a_{l}P_{l}(\cos \theta ),
\label{ParDev}
\end{equation}
where $x = \cos \theta$, $P_{l}(x)$ is the Legendre
polynomial of degree $l$,  $x \ne 1$, and $a_{l}$ is given 
in terms of the scattering matrix element $S_l$ by
\begin{equation}
a_l= \frac{1}{2ik} (2 l+ 1) S_l,
\label{ParAmp}
\end{equation}
where $k$ is the wavenumber.

To obtain the improved NF subamplitudes, one substitutes
the usual factor $S_l -1$ with $S_l$ 
on the r.h.s. of (\ref{ParAmp}). 
The dropped term ensured the convergence
of (\ref{ParDev}) for scattering by short range
potentials, for which $S_l \rightarrow 1$ exponentially for
$l \rightarrow \infty$ (\cite{DEA65}, p. 82).
In this case, having omitted a term 
$\propto \delta(1- {x})$, where $\delta$ indicates the Dirac
distribution (e.g. see \cite{BRI85}, p. 52), the sum
in (\ref{ParDev}) is defined only in a distributional
sense.
In the presence of a long range Coulomb term in the 
potential the dropped 1 is not relevant for convergence.
With or without the 1, the sum in (\ref{ParDev}) 
is convergent only in a distributional sense. 
In \cite{TAY74,GAR79,GES80,GOO80,QIO00}, and in references 
therein, one can find more or less recent discussions on the 
convergence of the Coulombic PWS, and of the 
different techniques (Pad\`e approximants, Abel summation, 
or different regularization procedures) solving the problem.

The improved NF subamplitudes are obtained by using
for $f(\theta)$, in place of (\ref{ParDev}), 
its resummed form
\begin{equation}
\label{ParDevRes}
f(\theta )=
\left ( 
\prod_{i=0}^{r} \frac{1}{1+\beta_i {x} }
\right )
\sum_{n=0}^{\infty} \alpha_{n}^{(r)}P_{n}({x} ),
\end{equation}
$r = 0, 1, 2, \ldots \, $, where
\begin{equation}
\label{RecRes}
\alpha_{n}^{(i)}= \beta_{i} \frac{n}{2 n-1} \alpha_{n-1}^{(i-1)}
+ \alpha_{n}^{(i-1)}
+ \beta_{i} \frac{n+1}{2 n+3} \alpha_{n+1}^{(i-1)},
\end{equation}
with $\beta_0 = 0$, $\alpha_n^{(0)}=a_n$, and 
$\alpha_{-1}^{(i)}=0$.
The resummed form (\ref{ParDevRes}) is an {\it exact mathematical 
identity} deriving from the recurrence property of the Legendre
polynomials. 
It holds for real or complex values of the 
resummation parameters $\beta_i \quad (i \ne 0$), 
restricted only by the condition 
$1+ \beta_i x \ne 0$, for $-1 \le x < 1$.
The integer index $r$ is the order of the resummation, and $r=0$
means no resummation of the original PWS.
In (\ref{ParDevRes}) we changed the index of the sum  
(\ref{ParDev}) (from $l$ to $n$) to remark that the 
index of the resummed Legendre polynomial series (LPS) 
in (\ref{ParDevRes}) has not, for $r \ne 0$,
the physical meaning of orbital quantum number,
differently from the index of the original PWS (\ref{ParDev}).
Similarly the terms $\alpha_{n}^{(i)}$ have not the 
physical meaning of partial wave amplitudes.
The usual YRW resummed form 
\cite{YEN54} for $f(\theta)$
is obtained by setting 
$\beta_i = -1$ ($i \ne 0$) in (\ref{ParDevRes}).

We note that for pure Coulomb scattering, for which
\begin{equation}
\label{CouA}
\alpha_{n}^{C(0)} \equiv a_n^C = \frac{1}{2ik} (2 n +1) 
\frac {\Gamma (n +1 +i \eta)}{\Gamma (n +1 -i \eta)}
\end{equation}
where $\eta$ is the Sommerfeld parameter, by using
(\ref{RecRes}) one obtains,  for large $n$ values, 
\begin{equation}
\label{CouR}
\alpha_{n}^{C(1)}= [1+ \beta_1 + O(n^{-2})]
\alpha_n^{C(0)}.
\end{equation}
This means that for $\beta_i \ne -1$ the asymptotic Coulombic
behavior of  $\alpha_{n}^{C(r)}$ does not depend, apart 
from a renormalization factor, on the resummation order $r$. 
On the other hand, given a resummed LPS of order $r$ (eventually
0), by applying an additional YRW ($\beta_{r+1}=-1$) 
resummation one obtains a convergent series for asymptotically Coulombic
$\alpha_{n}^{(r)}$.
Any successive YRW resummation improves the LPS convergence  
by a factor $O(n^{-2})$.

The improved NF subamplitudes are obtained by splitting 
in (\ref{ParDevRes}) the $P_n(x)$ into traveling angular
components
\begin{equation}
P_n({x})=Q_n^{(-)}({x})+Q_n^{(+)}({x}),
\label{EulDec}
\end{equation}
where (for $x \ne \pm 1$)
\begin{equation}  
\label{Qpm}
Q_n^{(\mp)}({x}) = 
\frac {1}{2}[ P_n({x}) \pm \frac {2i}{\pi}
Q_n({x}) ],
\end{equation}
with $Q_n({x})$ the Legendre function of the second 
kind of degree $n$.
By inserting (\ref{EulDec}) into (\ref{ParDevRes}), $f(\theta )$ 
is separated into the sum of two subamplitudes
\begin{equation}  
\label{NFFul}
f(\theta) = f^{(-)}_{\{\beta\}}(\theta) + f^{(+)}_{\{\beta\}}(\theta),
\end{equation}
with
\begin{equation}
\label{ParDevmp}
f^{(\mp )}_{\{\beta\}}(\theta )=
\left ( 
\prod_{i=0}^{r} \frac{1}{1+\beta_i {x} }
\right )
\sum_{n=0}^{\infty} \alpha_{n}^{(r)}Q_n^{(\mp)}({x}).
\end{equation}
In (\ref{ParDevmp}), with the subscript $\{\beta\}$ we indicate 
that the N ($f^{(-)}_{\{\beta\}}$) and F ($f^{(+)}_{\{\beta\}}$)
subamplitudes depend, differently from $f(\theta)$,  on 
the resummation order $r$ and parameters $\beta_i$.
%
This occurs because the resummed form of series (LFS) of linear 
combination of first and second kind Legendre functions, of 
integer degree, is different from (\ref{ParDevRes}).
In fact, let us indicate with
\begin{equation}
\label{ParDevE}
{\cal F}(\theta) = \sum_{n=0}^{\infty} d_n {\cal L}_n (x)
\end{equation}
a LFS in 
${\cal L}_n (x) = p P_n (x) + q Q_n(x)$, with $p$ and $q$
independent of $n$.
Owing to the property $n Q_{n-1} ({x}) \rightarrow 1$ as 
$n \rightarrow 0$ \cite{ANN81}, the resummed form of 
${\cal F}(x)$,  of order $s$ and 
parameters $\gamma_i$, is \cite{ANN02b}
\begin{eqnarray}
{\cal F}(\theta ) &=&
\left ( 
\prod_{i=0}^{s} \frac{1}{1+\gamma_i {x} }
\right )
\sum_{n=0}^{\infty} \delta_{n}^{(s)}{\cal L}_{n}({x} ) \nonumber \\
&+&q\sum_{i=0}^s \gamma_i \delta_0^{(i-1)} 
\prod_{j=0}^i \frac {1}{1+\gamma_j x}.
\label{ParDevResE}
\end{eqnarray}
Equation (\ref{ParDevResE}) is an {\it exact mathematical 
identity} extending the validity of (\ref{ParDevRes}) to
more general LFS, and it reduces to (\ref{ParDevRes}) for
LPS ($q=0$).
The conditions of validity of (\ref{ParDevResE}), and the 
recurrence relation for the resummed coefficients, are the 
same as those for (\ref{ParDevRes}), after substituting  
$r, \beta, \alpha$, and $a$ with $s, \gamma, \delta$, and $d$, 
respectively.

Because the $Q_n^{(\mp)} ({x})$ used to split
$P_n ({x})$ in (\ref{EulDec}) are a 
particular case of the more general ${\cal L}_n ({x})$
(with $p= 1/2$, and $q = \pm  i / \pi$),
the presence of the last term in (\ref{ParDevResE}) is 
responsible for the dependence of 
$f^{(\mp )}_{\{\beta\}}(\theta )$ on $r$ and $\beta_i$.
%
The last term on the r.h.s. of (\ref{ParDevResE}) gives a 
contribution if the splitting
(\ref{EulDec}) is inserted in (\ref{ParDev}).
This contribution is absent if the splitting is inserted
in (\ref{ParDevRes}).
%

In Refs. {\cite{ANN01d,ANN02b}} it was observed that unphysical 
contributions, when appearing in the Fuller
NF subamplitudes ($r=0$ in (\ref{ParDevmp})), decrease  
by increasing $r$ in (\ref{ParDevmp})  
(the values $r=1$, and $2$ were tested), 
if the $\beta_i$ are selected to make 
null the coefficients 
$\alpha_0^{(r)},
\alpha_1^{(r)}, \ldots 
\alpha_{r-1}^{(r)}$
of the resummed LFS ($\alpha_0^{(1)}$, and $\alpha_{0,1}^{(2)}$ 
for the cases tested).
In this way one drops the contributions to the NF resummed 
subamplitudes from low $n$ values for which the 
splitting (\ref{EulDec}), though exact by construction, 
is not expected to be physically meaningful.
%

The $\alpha_n^{(r)}$ in (\ref{ParDevRes}) and (\ref{ParDevmp})
go asymptotically to a constant for short range
potentials, or are Coulombic in the presence of a Coulomb
term in the potential.
Because of this the corresponding LFS are not convergent in the
usual sense.
In Refs.  {\cite{ANN01d,ANN02b} the convergence was forced, 
and accelerated, by applying to the improved LFS
a final (EYRW) resummation (\ref{ParDevResE}), of order
$s \ge 1$,
with $d_n=\alpha_n^{(r)}$, $\gamma_i =-1$, and $i \ne 0$.
%
%
%
The final EYRW resummation ensures the numerical convergence of 
the LFS, with a convergence rate increasing with $s$.
The increased rate of convergence costs, however, 
the cancellation of significant digits (see \cite{ANN81} 
for details), and 
numerically the procedure may results 
not convenient or even impossible, using arithmetic with a 
fixed digit number.

These troubles can be avoided by investigating 
the properties of the
resummation identity (\ref{ParDevResE}) with $d_n$ equal 
to the pure Coulomb $a_n^C$ given by (\ref{CouA}).
In this case we explicitly know the
l.h.s. of (\ref{ParDevResE}) for the relevant $p$ and 
$q$ values.
In fact, if $p =1$ and $q=0$ it is
the Rutherford scattering amplitude $f_R(\theta)$, while 
for $p= 1/2$ and $q = \pm  i / \pi$ one
obtains the Fuller-Rutherford NF 
subamplitudes $f_{FR}^{(\mp)}(\theta)$ (\cite{FUL75} Eqs. 14 a, b).
Because (\ref{ParDevResE}) is exact it holds for arbitrary
$\gamma_i$, and therefore also for  
$\gamma_i=\beta_i$, with $\beta_i$ obtained by applying the improved
resummation method to the exact $S_l$.
With this choice the pure Coulomb 
resummed coefficients
$\alpha_n^{C(r)}$ asymptotically approach $\alpha_n^{(r)}$ 
as rapidly as the pure Coulomb $S$-matrix elements, $S_l^C $, 
approach $S_l$ in the usual optical potential calculations.

With the change of notation 
$f^{(0)} \equiv f$,
$f^{(\mp 1)} \equiv f_{\{\beta\}}^{(\mp)}$, 
$f_R^{(0)} \equiv f_R$,
$f_R^{(\mp 1)} \equiv f_{FR}^{(\mp)}$, 
${\cal L}_n^{(0)} \equiv P_n$, and
${\cal L}_n^{(\mp 1)} \equiv Q_n^{(\mp)}$,
by subtracting from (\ref{ParDevRes}), or (\ref{ParDevmp}),
the corresponding resummed forms (\ref{ParDevResE}) applied to
pure Coulomb scattering ($ s = r, \gamma_i=\beta_i,
\delta_n^{(s)}=\alpha_n^{C (r)}$ and $q=0,\mp 1$), 
one obtains the final result
\begin{eqnarray}
f^{(m)}(\theta ) &=&
\left ( 
\prod_{i=0}^{r} \frac{1}{1+\beta_i {x} }
\right )
\sum_{n=0}^{\infty} [\alpha_{n}^{(r)}-\alpha_{n}^{C(r)}]
{\cal L}_{n}^{(m)}({x} )  \nonumber\\
&+& f_R^{(m)}(\theta )
+ m \frac {i}{\pi} \sum_{i=0}^r \beta_i \alpha_0^{C(i-1)} 
\prod_{j=0}^i \frac {1}{1+\beta_j x},\nonumber \\
\label{EqFin}
\end{eqnarray}
with $m=0$ for the full amplitude and
$m=\mp 1$ for the NF subamplitudes. 
For $r=0$ and $m=0$, or $m=\mp 1$, Eq. (\ref{EqFin}) 
is the usual regularization procedure defining
the r.h.s of (\ref{ParDevRes}), or (\ref{ParDevmp}),
in the presence of a long range Coulomb term in the potential.
This procedure is based on adding the explicit 
expression of $f_R(\theta)$, or $f^{(\mp)}_{FR} (\theta)$, and 
subtracting its formal PWS, or LFS,  for 
the full amplitude (Ref. \cite{MES61}, p. 428), or the Fuller NF 
subamplitudes \cite{FUL75}. 
For $r \ge 1$, Eq. (\ref{EqFin}) is the generalization
of this regularization procedure to resummed forms of the
full amplitude, or NF subamplitudes.
%
%
The sum appearing in this term is as rapidly 
convergent as the usual sum with $r=0$.

\begin{figure}
\centering
\includegraphics[width = 8.6 cm]{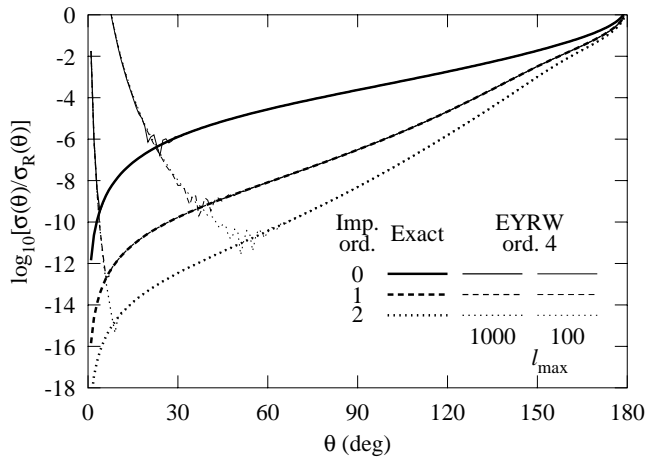}
\caption{\label{fig:Fig01} 
Different order ($r = 0, 1$ and $2$) improved F pure Coulomb  cross sections 
calculated with the exact expression (thick curves) and using an 
EYRW resummation of order 4, with 1000 (medium thickness curves) and 100 
(thin curves) partial waves.
}
\end{figure}

Before showing the effectiveness of our regularization 
procedure in a physically interesting case,
we show the difficulties met by the EYRW technique \cite{ANN81}
to ensure, and speed up, the convergence of improved, 
or not, LFS for pure Coulomb scattering.
In this case $a_n \equiv a_n^C$, and
the LFS on the 
r.h.s. of (\ref{EqFin}) is identically null, with arbitrary
choice of $\beta_i$.
For $r=0$, Eq. (\ref{EqFin}) trivially states that the scattering 
amplitude ($m=0$) is the Rutherford amplitude, and 
the NF subamplitudes ($m=\mp 1$) are the usual Fuller-Rutherford ones.
For $r > 0$, by choosing $\beta_i$ accordingly with
the improved resummation method, Eq. (\ref{EqFin}) gives the
explicit expression of the improved NF subamplitudes ($m=\mp 1$) 
in term of the usual Fuller-Rutherford ones, and of simple functions 
depending on $\beta_i$ and $\alpha_0^{C(i-1)}$.
For simplicity we will name {\it exact} this explicit 
expression for pure Coulomb improved NF subamplitudes.
 
In Fig. 1 the thick curves show the ratio to the 
Rutherford cross section, $\sigma_R(\theta)$, of the exact 
pure Coulomb improved F cross sections, of order $r = 0, 1,$ and 2 
($r = 0$ meaning the original
Fuller method).
In the same figure the thin curves show the
F cross sections obtained by forcing, and accelerating, the convergence 
of (\ref{ParDevmp}) with an additional EYRW 
resummation of order $s = 4$, and 
fixing the maximum number of the summed partial waves to 
$l_{\text {max}} = 100$ and $1000$.
The results were obtained with $\eta =10$, which is a 
typical value of the Sommerfeld parameter for heavy-ion scattering.
For this $\eta$ value the improved resummation parameters are 
$\beta_1= 0.9802 + 0.1980 i$ (for $r=1$), 
$\beta_1=1.0072 + 0.1166 i$ and $\beta_2 = 0.7804 + 0.6052 i$
(for $r=2$).
Figure 1 shows that the final EYRW resummation 
\cite{ANN81} ensures
the convergence of the LFS (\ref{ParDevmp}),
but the convergence rate is low.
For $\theta \lesssim 5^\circ$ a numerically satisfactory result
is not obtained even with $l_{\text {max}} = 1000$.
By fixing $l_{\text {max}}$ and the final resummation order, 
the angle at with the truncated LFS disagrees with the 
exact result increases with the improved 
resummation order.

\begin{figure}
\centering
\includegraphics[width = 8.6 cm]{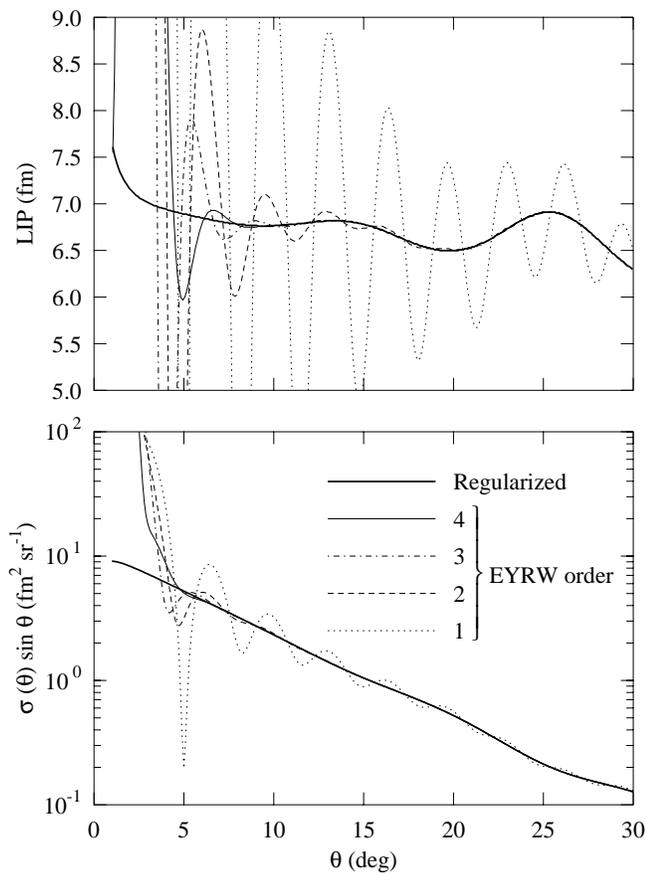}
\caption{\label{fig:Fig02} 
First order ($r = 1$) improved F cross section (lower panel)
and LIP (upper panel) for the $^{16}$O + $^{16}$O case.
The calculations were done using 150 partial waves with 
our regularization procedure (thick curves) and with different 
final EYRW resummation orders (thin curves).
}
\end{figure}

Figure 1 also shows that the improved resummation method
reduces, particularly at forward angles, the unphysical F 
contribution present in the original Fuller NF method.
However it does not suppress it, and is ineffective at 
$\theta \approx 180^\circ$.
This is an insurmountable difficulty connected with the
NF splitting (\ref{EulDec}), mathematically continuing 
(at $\theta = 180^\circ$) the N subamplitude into a F one,
or vice versa.
This also in absence of physically meaningful 
subamplitudes justifying this continuation.
%
%
In these situations the only practical suggestion we can give 
is to not take seriously the NF subamplitudes at 
$\theta \approx 180^\circ$, if in a neighbourood of this 
angle the cross section and the LIP of the full amplitude 
have a non oscillatory behavior, suggesting the dominance of 
a {\it single side} (positive LIP for F and negative for N) 
contribution. 
We remember that in \cite{ANN01d} the LIP (local impact
parameter) is defined as the derivative 
of the argument of the scattering amplitude with respect to the 
scattering angle, named LAM (local angular momentum) by Fuller\cite{FUL75}, 
divided by the wavenumber $k$.

As a second example of the effectiveness of our regularization
procedure, we consider the first order improved F cross 
section and LIP of the phenomenological optical potential WS2, used 
to fit \cite{KHO00} the $^{16}$O + $^{16}$O elastic cross section
at $E_{\text {lab}} = 145$ MeV. 
The improved resummation parameter is in this case $\beta_1=
-.9997 -0.0798 i$ \cite{ANN02b}.
The upper panel of Fig. 2 shows, for $\theta < 30^\circ$ and 
$l_{\text{max}} = 150$, the F LIP 
calculated using 
our regularization procedure 
(thick curve) and  different order (thin lines) 
EYRW resummations \cite{ANN81}.
The lower panel shows the corresponding F cross sections.
Symmetrization effects were ignored.

Note that 150 partial waves are more than really 
necessary to obtain reliable scattering amplitudes using 
our regularization procedure.
Using an EYRW resummation of order 1 (thin dotted curves), this 
partial wave number is not sufficient to obtain a
satisfactory result.
By increasing the EYRW resummation order it 
decreases the angular width of the region where the thin curves
differ from the corresponding thick ones.
However, for $\theta \lesssim 5 ^\circ$, the 150 partial
waves used are not enough, even using a fourth order final 
EYRW resummation.
 
These results show, in practical examples, that EYRW 
resummed LFS for asymptotically Coulombic $S_l$ are convergent,
with a convergence rate increasing by increasing the
resummation order.
Compared with the extension here given of the usual
regularization procedure for asymptotically Coulombic $S_l$ 
the EYRW resummation technique effectiveness is, however, 
computationally poor.
The regularization procedure here described can be 
easily extended to make rapidly convergent the LFS
in (\ref{ParDevRes}) and (\ref{ParDevmp}) for 
scattering by short range potentials.
In these cases, however, also an additional 
first order EYRW resummation makes the LFS convergent 
with the same rapidity, and there is no practical advantage
in using a different procedure.

\begin{acknowledgments}
The author is indebted to J.~N.~L. Connor for stimulating and
helpfull discussions.
\end{acknowledgments}


\end{document}